\documentclass[fleqn,a4paper,11pt]{article} 
\usepackage{graphicx,xcolor,physics,fullpage,amsmath,amssymb,amsthm,eufrak}
\usepackage[title]{appendix}
\usepackage[numbers,sort&compress]{natbib}
\usepackage[hidelinks,colorlinks=true,linkcolor=blue,citecolor=blue,urlcolor=blue]{hyperref}
\theoremstyle{remark}

\newcommand{\eq}[2]{\begin{gather} #2  \label{#1} \end{gather}}
\newcommand{\nn}{\nonumber}
\renewcommand{\braket}[2]{\left\langle #1, #2 \right\rangle}

\title{\bf Energy Method and Stability of Shear Flows:\\ an Elementary Tutorial}
\author{\bf Antonio Barletta$^{(1)}$\orcidlink{0000-0002-6994-5585}, Giuseppe Mulone$^{(2)}$\orcidlink{0000-0002-9136-9748}
\\[4pt]
$^{(1)}${\small Department of Industrial Engineering, University of Bologna,} \\
{\small Viale Risorgimento 2, 40136 Bologna, Italy,}\\ 
\small{\href{mailto:antonio.barletta@unibo.it}{antonio.barletta@unibo.it} (corresponding author)}\\[2pt]
$^{(2)}${\small Department of Mathematics and Computer Science, University of Catania\footnote{Retired}~,} \\
{\small Viale Andrea Doria 6, 95125 Catania, Italy.}\\ 
\small{\href{mailto:giuseppe.mulone@unict.it}{giuseppe.mulone@unict.it}}
}
\date{} 
         
\usepackage{orcidlink}         

\begin{document}
\maketitle

\begin{abstract}\noindent
This paper provides a pedagogical introduction to the classical nonlinear stability analysis of the plane Poiseuille and Couette flows. The whole procedure is kept as simple as possible by presenting all the logical steps involved in the application of the energy method and leading to the Euler-Lagrange equations. Then, the eigenvalue problems needed for the evaluation of the nonlinear energy threshold of the Reynolds number for stability are formulated for transverse modes and for longitudinal modes. Such formulations involve the streamfunction and, in the case of longitudinal modes, also the streamwise component of velocity.  
An accurate numerical solution of the eigenvalue problems, based on Galerkin's method of weighted residuals with the test functions expressed in terms of Chebyshev polynomials, is discussed in details. The numerical codes developed for the software {\sl Mathematica 14} ({\small \copyright}~Wolfram Research, Inc.) are also presented. A critical analysis of the obtained results is finally proposed.
\end{abstract}

\section{Introduction}
The hydrodynamic instability of shear flows is a cornerstone topic in fluid dynamics. A viscous fluid flow may become unstable and, eventually, turbulent due to increasingly intense velocity gradients. This phenomenon is crucial to understand manifold natural or industrial processes. One may mention atmospheric circulation and oceanic currents, but also engineering applications such as the design of aircrafts or pipelines for oil or gas transport. There are many textbooks describing the mathematical modelling of shear flows and their stability/instability transition, by employing either a linear analysis based on the dynamics of infinitesimal amplitude perturbations or a nonlinear analysis where the evolution of perturbations having a finite amplitude is investigated. Among them, we mention the classical treatises by \citet{joseph1976stability1}, by \citet{schmid2001stability} and by \citet{drazin2004hydrodynamic}. 

There are many mathematical methods that may be employed to gather information on the stability/instability of shear flows. On one hand, the linear instability analysis monitors the time evolution of perturbations acting upon the basic shear flow and having an infinitesimally small amplitude. Such a type of analysis leads to the detection of the lowest parametric threshold above which stability is ruled out. On the other hand, the energy method is appropriate to detect the highest parametric threshold below which no instability (linear or nonlinear) may happen. In cases where the two thresholds are distinct, the linear method and the energy method cannot provide directly any information for the parametric region in between such thresholds \citep{joseph1976stability1,schmid2001stability}. There is an important exception to this general rule which is grounded on the transient growth analysis. Indeed, a linearly stable basic flow might show a transient growth when the evolution of the perturbations is described by a non-normal linear operator, so that its eigenvectors are non-orthogonal \cite{schmid2007nonmodal}. This is a case where a purely linear analysis may provide insights into a subcritical instability. However, such a use of the linear analysis will not be developed further in this paper, as our purpose is a pedagogical introduction to the energy method in all its aspects including the numerical implementation and the coding via a weighted residual method based on Chebyshev polynomial series. Furthermore, the linear analysis of the transition to instability will not be discussed here as its details together with the coding of a numerical solver for its exploitation have been already presented in a previous paper \cite{barletta2024alternative}. 

The focus of our paper is on the nonlinear stability analysis of the plane Couette and Poiseuille flows in a plane parallel channel as carried out with the energy method. The main purpose is not to convey new result, but to provide a step-by-step primer to all the assumptions, methodology and numerical solution, including all the coding needed to obtain numerical data via the software {\sl Mathematica 14} ({\small \copyright}~Wolfram Research, Inc.). This pedagogical aim will be accomplished relying on the well-settled knowledge on this subject resulting from the less recent literature, as well as on the more recent results provided in the papers by \citet{xiong2019conjecture}, \citet{fuentes2022global}, \citet{nagy2022enstrophy} and by \citet{cui2024review}. Further recent research on this topic features the critical analysis of the role played by transverse and longitudinal modes in the solution of the Euler-Lagrange equation and the evaluation of the nonlinear stability upper bound as defined by the energy method 
\citep{PhysRevE.100.013113,FALSAPERLA202293,Mulone2024}. A gap in the huge literature available on this subject is the detailed description of all the logical-deductive steps needed to formulate the energy method analysis of the nonlinear stability of Couette and Poiseuille flows in a plane channel, leading to the Euler-Lagrange equations and the eigenvalue problem achieved when the optimisation of the energy bound to stability is performed. We mention that also the coding details of the numerical solution for such an eigenvalue problem are hardly exploited in the existing literature. The latter feature is shared with a couple of recent papers \citep{arnone2023chebyshev,barletta2024alternative}, where the latter provides a detailed description of the $\tau$-method and its coding.

This paper is structured as follows. Section~\ref{goeq} provides a detailed outline of the Couette and Poiseuille flows in a parallel plane channel, together with their nonlinear stability analysis grounded on the energy method. Section~\ref{enme} shows the subtleties of the optimisation process in the definition of the upper bound to stability as defined by the energy method, leading eventually to the statement of  the Euler-Lagrange equations. Section~\ref{orr} provides the formulation of the stability eigenvalue problem based on the Euler-Lagrange equations when specialised to the transverse modes, or T-modes. Section~\ref{joseph} is the dual of Section~\ref{orr} when longitudinal modes, or L-modes, are employed instead of T-modes. The numerical solution tool and its coding, developed through the weighted residual method and Chebyshev polynomials expansions, are presented for the software {\sl Mathematica 14} in the Appendices~\ref{AppA} and \ref{AppB}. Section~\ref{dire} provides a thorough and critical description of the numerical results obtained with hints for their physical interpretation.

\section{Governing equations}\label{goeq}
The study focusses on the plane Poiseuille and Couette flows. For the sake of simplicity, we will start from the dimensionless governing equations where the reference scales employed are:
\begin{itemize}
\item Cartesian coordinates are scaled by the channel half-width;
\item Velocity components are scaled by the maximum velocity of either Poiseuille flow or Couette flow;
\item Pressure is scaled by the product between the density and the square reference velocity.
\end{itemize}
Consistently, time is scaled by the ratio between the reference length and the reference velocity, while the Reynolds number is defined by dividing the product of the reference velocity and the reference length by the kinematic viscosity.

The dimensionless governing equations for the shear flows in a channel with $(x,y) \in \mathbb{R}$ and $z \in[-1,1]$ express the local mass balance and the local momentum balance for incompressible flow,
\eq{1}{
\pdv{U_j}{x_j} = 0, \nn\\
\pdv{U_i}{t} + U_j \pdv{U_i}{x_j} = - \pdv{P}{x_i} + \frac{1}{Re} \laplacian{U_i},
}
where $\vb{U}$ with $i$th component $U_i$ is the velocity field, $P$ is the pressure field, $x_i$ and $t$ are the $i$the coordinate and time, while $Re$ is the Reynolds number. Einstein's notation for implicit sums over repeated indices is used. 

\subsection{Basic shear flows}
Stationary basic solutions of \eqref{1} can be found having the form
\eq{2}{
U = f(z) \qc V = 0 \qc W = 0 \qc \pdv{P}{x} = - \frac{2\sigma}{Re} \qc \pdv{P}{y} = 0 \qc \pdv{P}{z} = 0 ,
}
where $\sigma$ is a constant, $(U,V,W)$ denotes the Cartesian components of the velocity $U_i$ and $f(z)$ is determined so that \eqref{1} is satisfied, namely
\eq{3}{
f''(z) = - 2 \sigma .
}
One can devise two focus cases,
\eq{4}{
f(z) = 1 - z^2 \qquad \sigma = 1 \quad \text{(Poiseuille flow)}, \nn\\
f(z) = z \qquad \sigma = 0 \quad \text{(Couette flow)}.
}
\subsection{Dynamics of perturbations}
One can perturb the basic state given by \eqref{2} and \eqref{4}. Let us define
\eq{5}{
U = f(z) + u \qc V = v \qc W = w \qc \pdv{P}{x} = - \frac{2\sigma}{Re} + \pdv{p}{x} \qc \pdv{P}{y} = \pdv{p}{y} \qc \pdv{P}{z} = \pdv{p}{z} .
}
By substituting \eqref{5} in \eqref{1}, one obtains
\eq{6}{
\pdv{u_j}{x_j} = 0, \nn\\
\pdv{u}{t} + f(z) \pdv{u}{x} + w f'(z) + u_j \pdv{u}{x_j} = - \pdv{p}{x} + \frac{1}{Re} \laplacian{u} , \nn\\
\pdv{v}{t} + f(z) \pdv{v}{x} + u_j \pdv{v}{x_j} = - \pdv{p}{y} + \frac{1}{Re} \laplacian{v} , \nn\\
\pdv{w}{t} + f(z) \pdv{w}{x} + u_j \pdv{w}{x_j} = - \pdv{p}{z} + \frac{1}{Re} \laplacian{w} , 
}
where $u_i$ is the $i$th component of the velocity perturbation $(u,v,w)$. The boundary and initial conditions are
\eq{7}{
u_i (x,y, \pm1,t) = 0 \qc u_i (x,y,z,0) = \tilde{u}_i(x,y,z) ,
}
where vector $\tilde{u}_i$ is solenoidal and vanishes at $z = \pm 1$.

We further assume that the perturbations are periodic both in the $x$ and in the $y$ directions with periods $2\pi/\alpha_x$ and $2\pi/\alpha_y$, respectively, where $\alpha_x$ and $\alpha_y$ are the wavenumbers.

We introduce the shortcut notation for the scalar product and for the norm,
\eq{8}{
\bra{g} \ket{h}  = \int\limits_{\Omega} g\, h\ \dd^3 x \qc \norm{g}^2 =  \int\limits_{\Omega} |g|^2\ \dd^3 x , \quad \Omega = \left[ 0, \frac{2\pi}{\alpha_x} \right] \times \left[ 0, \frac{2\pi}{\alpha_y} \right] \times [-1,1],
}
with $g, h \in {\cal L}^2(\Omega)$.

We multiply the second \eqref{6} by $u$, the third \eqref{6} by $v$ and the fourth \eqref{6} by $w$. Then, we integrate over $\Omega$,
\eq{9}{
\frac{1}{2} \dv{}{t} \norm{u}|^2 + \bra{u} \ket{f \pdv{u}{x}} + \bra{u} \ket{w f'} + \bra{u} \ket{u_j \pdv{u}{x_j}} + \bra{u} \ket{\pdv{p}{x}} - \frac{1}{Re} \bra{u} \ket{\laplacian{u}} = 0, \nn\\
\frac{1}{2} \dv{}{t} \norm{v}^2 + \bra{v} \ket{f \pdv{v}{x}} + \bra{v} \ket{u_j \pdv{v}{x_j}} + \bra{v} \ket{\pdv{p}{y}} - \frac{1}{Re} \bra{v} \ket{\laplacian{v}} = 0, \nn\\
\frac{1}{2} \dv{}{t} \norm{w}^2 + \bra{w} \ket{f \pdv{w}{x}} + \bra{w} \ket{u_j \pdv{w}{x_j}} + \bra{w} \ket{\pdv{p}{z}} - \frac{1}{Re} \bra{w} \ket{\laplacian{w}} = 0.
}
Let us sum the three equations \eqref{9}
\eq{10}{
\frac{1}{2} \dv{}{t} \norm{\vb{u}}^2 + \bra{u_j} \ket{f \pdv{u_j}{x}} + \bra{u} \ket{w f'} + \bra{u_i} \ket{u_j \pdv{u_i}{x_j}} \nn\\
\hspace{5cm}+ \bra{u_j} \ket{\pdv{p}{x_j}} - \frac{1}{Re} \bra{u_j} \ket{\laplacian{u_j}} = 0, 
}
where
\eq{11}{
\norm{\vb{u}}^2 = \norm{u}^2 + \norm{v}^2 + \norm{w}^2 .
}

\subsubsection*{Term $\displaystyle \bra{u_j} \ket{f \pdv{u_j}{x}}$}
The second term on the left hand side of \eqref{10} can be rewritten as
\eq{12}{
\bra{u_j} \ket{f \pdv{u_j}{x}} = \int\limits_{\Omega} f(z) u_j \pdv{u_j}{x} \ \dd^3 x = \frac{1}{2} \int\limits_{\Omega} f(z) \pdv{|\vb{u}|^2}{x} \ \dd^3 x \nn\\
\hspace{5cm} = \frac{1}{2} \left. \int\limits_{[0,2\pi/\alpha_y]\times[-1,1]} f(z) |\vb{u}|^2 \dd y\, \dd z \right|_{x=2\pi/\alpha_x} 
 \nn\\
\hspace{7cm}- \frac{1}{2} \left. \int\limits_{[0,2\pi/\alpha_y]\times[-1,1]} f(z) |\vb{u}|^2 \dd y\, \dd z \right|_{x=0} = 0,
}
where 
the periodicity conditions at $x=0$ and $x=2\pi/\alpha_x$ have been employed. 

\subsubsection*{Term $\displaystyle \bra{u_i} \ket{u_j \pdv{u_i}{x_j}}$}
We can write
\eq{14}{
\bra{u_i} \ket{u_j \pdv{u_i}{x_j}} = \int\limits_{\Omega} u_i u_j \pdv{u_i}{x_j} \ \dd^3 x = \frac{1}{2} \int\limits_{\Omega} u_j \pdv{|\vb{u}|^2}{x_j} \ \dd^3 x = \frac{1}{2} \int\limits_{\partial\Omega} |\vb{u}|^2 \vb{u} \vdot \vb{n} \ \dd^2 x \nn\\
\hspace{5cm} - \frac{1}{2} \int\limits_{\Omega} |\vb{u}|^2 \pdv{u_j}{x_j} \ \dd^3 x = 0 ,
}
where $\partial\Omega$ is the boundary of $\Omega$ with $\vb{n}$ the unit outward normal to $\partial\Omega$, Gauss' theorem has been used together with \eqref{7} and the periodicity conditions over $x$ and $y$. Finally, the first \eqref{6} has been invoked.

\subsubsection*{Term $\displaystyle \bra{u_j} \ket{\pdv{p}{x_j}}$}
This term can be rewritten as
\eq{15}{
\bra{u_j} \ket{\pdv{p}{x_j}} = \int\limits_{\Omega} u_j \pdv{p}{x_j}\ \dd^3 x = \int\limits_{\partial\Omega} p\, \vb{u} \vdot \vb{n}\ \dd^2 x - \int\limits_{\Omega} p \pdv{u_j}{x_j}\ \dd^3 x = 0,
}
where Gauss' theorem together with \eqref{7}, the periodicity conditions over $x$ and $y$ and the first \eqref{6} have been employed. 

\subsubsection*{Term $\displaystyle \bra{u_j} \ket{\laplacian{u_j}}$}
By employing again Gauss' theorem together with \eqref{7}, the periodicity conditions over $x$ and $y$ and the first \eqref{6}, one has
\eq{16}{
\bra{u_j} \ket{\laplacian{u_j}} = \int\limits_{\Omega} u_j \, \laplacian{u_j} \ \dd^3 x = \int\limits_{\partial\Omega} u_j  n_i \pdv{u_j}{x_i} \ \dd^2 x - \int\limits_{\Omega} \pdv{u_j}{x_i} \pdv{u_j}{x_i} \ \dd^3 x \nn\\
\hspace{5cm}= - \bra{\pdv{u_j}{x_i}} \ket{\pdv{u_j}{x_i}} = - \norm{ \grad{\vb{u}}}^2.
}
The last equality of \eqref{16} is to be intended as a notational definition. An alternative expression is
\eq{17}{
\norm{ \grad{\vb{u}}}^2 = \norm{ \grad{u}}^2 + \norm{ \grad{v}}^2 + \norm{ \grad{w}}^2 .
}

\section{Energy method}\label{enme}
We are now ready for rewriting \eqref{10} as the Reynolds-Orr equation,
\eq{18}{
\frac{1}{2} \dv{}{t} \norm{\vb{u}}^2 + \bra{u} \ket{w f'} + \frac{1}{Re} \norm{ \grad{\vb{u}}}^2 = 0.
}
Furthermore, we introduce the dimensionless kinetic energy, hereafter called energy for brevity,
\eq{19}{
E(t) = \frac{1}{2} \norm{\vb{u}}^2 .
}
Thus, \eqref{18} can be rewritten as
\eq{20}{
\dv{E}{t} = - \bra{u} \ket{w f'} - \frac{1}{Re} \norm{ \grad{\vb{u}}}^2 = \left(\frac{- \bra{u} \ket{w f'}}{\norm{ \grad{\vb{u}}}^2} - \frac{1}{Re} \right) \norm{ \grad{\vb{u}}}^2 .
}
We define 
\eq{21}{
m = \max_{\cal S} \left(\frac{- \bra{u} \ket{w f'}}{\norm{ \grad{\vb{u}}}^2} \right) = \max_{\cal S} \left(\frac{- \bra{u} \ket{w f'}}{\norm{ \grad{u}}^2 + \norm{ \grad{v}}^2 + \norm{ \grad{w}}^2} \right) ,
}
with ${\cal S}$ a subspace of the Sobolev space ${\cal H}^1(\Omega)$. The elements of ${\cal S}$ are the kinematically admissible velocity fields, $\vb{u} = (u,v,w)$, namely those satisfying the conditions $\vb{u}=0$ at $z=\pm1$, the periodicity conditions along $x$ and $y$, and the constraint of zero divergence given by the first \eqref{6}.
As a consequence of \eqref{20} and \eqref{21}, one obtains 
\eq{22}{
\dv{E}{t} \le \left(m - \frac{1}{Re} \right) \norm{ \grad{\vb{u}}}^2 .
}
From \eqref{22}, it is evident that the energy is monotonically decreasing with time when
\eq{23}{
Re < \frac{1}{m} .
}
We now assume the validity of \eqref{23}, and we make use of the Poincar\'e inequality,
\eq{24}{
\norm{ \grad{\vb{u}}}^2 \ge \xi \, \norm{ \vb{u} }^2,
}
where $\xi$ is a positive constant. Thus, \eqref{22} and \eqref{24} yield
\eq{25}{
\dv{E}{t} \le \xi \left(m - \frac{1}{Re} \right) \norm{ \vb{u}}^2 ,
}
or equivalently
\eq{26}{
E(t) \le E(0) \exp\qty[2 \xi \qty(m - \frac{1}{Re}) t ] .
}
Inequality \eqref{26} shows that not only $E(t)$ is a monotonically decreasing function, but its asymptotic value for $t \to +\infty$ is zero. This result implies that \eqref{23} is a sufficient condition for stability. The energy threshold for stability is
\eq{27}{
Re_E = \frac{1}{m} .
}

\subsection{Properties of $m$}\label{propm}
On account of \eqref{21}, the value of $m$ is invariant under an overall scaling of the vector field $\vb{u}$, namely $m$ does not change if 
\eq{28}{
\vb{u} \to C\vb{u}, 
}
for every constant $C \in \mathbb{R}$. This means that $m$ is not influenced by the overall scale of the perturbations. Roughly speaking, one can consider small or large amplitude perturbations with no effects on the value of $m$ and hence of the threshold Reynolds number, $Re_E$. Another way to formulate such a result is that the nonlinearity of \eqref{6} does not impact on the value of $m$. Should one have neglected the nonlinear terms in \eqref{6}, the evaluation of $m$ would not have been any different.

\subsection{Evaluation of $m$}
Let $\vb{\hat{u}}=\qty(\hat{u}, \hat{v}, \hat{w})$ be the element of $\cal S$ yielding the maximum defined by \eqref{21}, namely
\eq{29}{
m = \frac{- \bra{\hat{u}} \ket{\hat{w} f'}}{\norm{ \grad{\vb{\hat{u}}}}^2} .
}
Let us define the functions
\eq{30}{
I(\epsilon) = - \braket{\hat{u} + \epsilon \, u_1}{\qty(\hat{w} + \epsilon\, w_1) f'} = 
- \braket{\hat{u}}{\hat{w} f'} - \epsilon \Big( \braket{\hat{u}}{w_1 f'} + \braket{u_1}{\hat{w} f'} \Big) \nn\\
\hspace{1.5cm}-\, \epsilon^2 \braket{u_1}{w_1 f'},  \nn\\
D(\epsilon) = \norm{ \grad{\qty(\hat{u} + \epsilon \, u_1)}}^2 + \norm{ \grad{\qty(\hat{v} + \epsilon \, v_1)}}^2 + \norm{ \grad{\qty(\hat{w} + \epsilon \, w_1)}}^2 \nn\\
\hspace{1.5cm}= \norm{\grad{\hat{u}}}^2 + \norm{\grad{\hat{v}}}^2 + \norm{\grad{\hat{w}}}^2 + 2 \epsilon \Big( \braket{\grad{\hat{u}}}{\grad{u_1}} + \braket{\grad{\hat{v}}}{\grad{v_1}} + \braket{\grad{\hat{w}}}{\grad{w_1}} \Big) \nn\\
\hspace{3cm}+\ \epsilon^2 \Big( \norm{\grad{u_1}}^2 + \norm{\grad{v_1}}^2 + \norm{\grad{w_1}}^2 \Big),
}
with $\epsilon \in \mathbb{R}$ and $\vb{u}_1 = (u_1, v_1, w_1) \in {\cal S}$. If $\vb{\hat{u}}$ satisfies \eqref{29}, then $\epsilon = 0$ must be a stationary point of the ratio $I(\epsilon)/D(\epsilon)$, namely
\eq{31}{
\left.\dv{}{\epsilon} \qty[\frac{I(\epsilon)}{D(\epsilon)}]\right|_{\epsilon=0} = 0 .
}
In other words, one has
\eq{32}{
\frac{1}{D(0)} \left. \dv{I(\epsilon)}{\epsilon} \right|_{\epsilon=0} - \frac{I(0)}{[D(0)]^2} \left. \dv{D(\epsilon)}{\epsilon} \right|_{\epsilon=0} = 0 .
}
We multiply \eqref{32} by $D(0)$ and recall that $I(0)/D(0) = m$. Thus, \eqref{32} can be rewritten as
\eq{33}{
\left. \dv{I(\epsilon)}{\epsilon} \right|_{\epsilon=0} - m \left. \dv{D(\epsilon)}{\epsilon} \right|_{\epsilon=0} = 0 , \nn\\
\braket{\hat{u} f'}{w_1} + \braket{\hat{w} f'}{u_1} + 2 m \Big( \braket{\grad{\hat{u}}}{\grad{u_1}} + \braket{\grad{\hat{v}}}{\grad{v_1}} + \braket{\grad{\hat{w}}}{\grad{w_1}} \Big) = 0 .
}
If one considers the terms multiplying $2m$ in \eqref{33} they can be easily rewritten by employing Gauss' theorem and the boundary conditions on $\partial \Omega$. In fact, we have
\eq{34}{
\braket{\grad{\hat{u}}}{\grad{u_1}} = \int\limits_{\partial \Omega} u_1\, \vb{n} \vdot \grad{\hat{u}}\ \dd^2 x - \braket{\laplacian{\hat{u}}}{u_1} = - \braket{\laplacian{\hat{u}}}{u_1} .
}
Just the same reasoning can be applied to $\braket{\grad{\hat{v}}}{\grad{v_1}}$ and $\braket{\grad{\hat{w}}}{\grad{w_1}}$. Thus, \eqref{33} now reads
\eq{35}{
\braket{\hat{u} f'}{w_1} + \braket{\hat{w} f'}{u_1} - 2 m \Big( \braket{\laplacian{\hat{u}}}{u_1} + \braket{\laplacian{\hat{v}}}{v_1} + \braket{\laplacian{\hat{w}}}{w_1} \Big) = 0 , \nn\\
\braket{\hat{w} f' - 2 m\, \laplacian{\hat{u}}}{u_1} + \braket{-\,2m \,\laplacian{\hat{v}}}{v_1} + \braket{\hat{u} f' - 2m\, \laplacian{\hat{w}}}{w_1} = 0.
}

\subsection{Lagrange multiplier}
On account of \eqref{35}, we have a vector $\vb{Y} = \qty(Y_x, Y_y, Y_z)$ with
\eq{36}{
Y_x = \hat{w} f' - 2 m\, \laplacian{\hat{u}} \qc Y_y = -\,2m \,\laplacian{\hat{v}} \qc Y_z = \hat{u} f' - 2m\, \laplacian{\hat{w}} , 
}
such that 
\eq{37}{
\braket{Y_x}{u_1} + \braket{Y_y}{v_1} + \braket{Y_z}{w_1} = 0.
}
It is easily verified that, with any differentiable scalar field $\lambda$, $\vb{Y} = - \grad{\lambda}$ satisfies \eqref{37}. The scalar field $\lambda$ is the {\em Lagrange multiplier}. In fact, $\vb{u}_1 \in {\cal S}$ must satisfy the zero velocity and periodicity conditions at the boundary $\partial \Omega$, so that Gauss' theorem allows one to rewrite the left hand side of \eqref{37} as
\eq{38}{
- \braket{\lambda}{\pdv{u_1}{x}} - \braket{\lambda}{\pdv{v_1}{y}} - \braket{\lambda}{\pdv{w_1}{z}} = 
- \braket{\lambda}{\pdv{u_1}{x} + \pdv{v_1}{y} + \pdv{w_1}{z}} = 0 .
}
where the equality to zero is a consequence of $\vb{u}_1$ being solenoidal. As it is shown in Appendix B2 of the book by \citet{joseph1976stability1}, the second fundamental lemma of the calculus of variations implies that $\vb{Y} = - \grad{\lambda}$ is not only a sufficient condition, but also a necessary condition for \eqref{37} to be satisfied.

On account of \eqref{36}, together with the necessary and sufficient condition for \eqref{37}, namely $\vb{Y} = - \grad{\lambda}$, we can write the system of Euler-Lagrange equations,
\eq{39}{
\begin{cases}
\displaystyle
2 m\, \laplacian{\hat{u}} - \hat{w} f' = \pdv{\lambda}{x},\\
\displaystyle
2m \,\laplacian{\hat{v}} = \pdv{\lambda}{y},\\
\displaystyle
2m\, \laplacian{\hat{w}} - \hat{u} f' = \pdv{\lambda}{z} .
\end{cases}
}
On account of \eqref{7}, $\vb{\hat{u}}$ satisfies the boundary conditions
\eq{40}{
\hat{u} (x,y, \pm1,t) = 0 \qc \hat{v} (x,y, \pm1,t) = 0 \qc \hat{w} (x,y, \pm1,t) = 0 ,
}
while the initial condition \eqref{7} is  not relevant as system \eqref{39} is time-independent. This means that, without any loss of generality, solutions can be sought which have no explicit dependence on time.

\section{T-modes eigenvalue problem}\label{orr}
Let us assume that $\vb{\hat{u}}$ and the Lagrange multiplier $\lambda$ are independent of $y$. Hereafter, such a case will be termed {\em transverse modes} and abbreviated with T-modes. Then, the zero-divergence condition on $\vb{\hat{u}}$ can be expressed as,
\eq{41}{
\pdv{\hat{u}}{x} + \pdv{\hat{w}}{z} = 0,
}
which is identically satisfied on defining a streamfunction $\psi$ such that
\eq{42}{
\hat{u} = \pdv{\psi}{z} \qc \hat{w} = - \pdv{\psi}{x} .
}
By employing \eqref{42}, the first and the third \eqref{39} yield
\eq{43}{
\begin{cases}
\displaystyle
2 m\, \laplacian{\pdv{\psi}{z}} + \pdv{\psi}{x}\, f' = \pdv{\lambda}{x},\\[12pt]
\displaystyle
2m\, \laplacian{\pdv{\psi}{x}} + \pdv{\psi}{z}\, f' = - \pdv{\lambda}{z} .
\end{cases}
}
By deriving the first \eqref{43} with respect to $z$, the second \eqref{43} with respect to $x$ and summing the two resulting equations, we encompass the dependence on the Lagrange multiplier $\lambda$ and we get 
\eq{44}{
2 m\, \nabla^4{\psi} + 2 \pdv[2]{\psi}{x}{z}\, f' + \pdv{\psi}{x}\, f''= 0 ,
}
where $\nabla^4$ denotes the Laplacian of the Laplacian, $\nabla^4 = \laplacian{\laplacian}$.

\begin{figure}[t]
\centering
\includegraphics[width=0.5\textwidth]{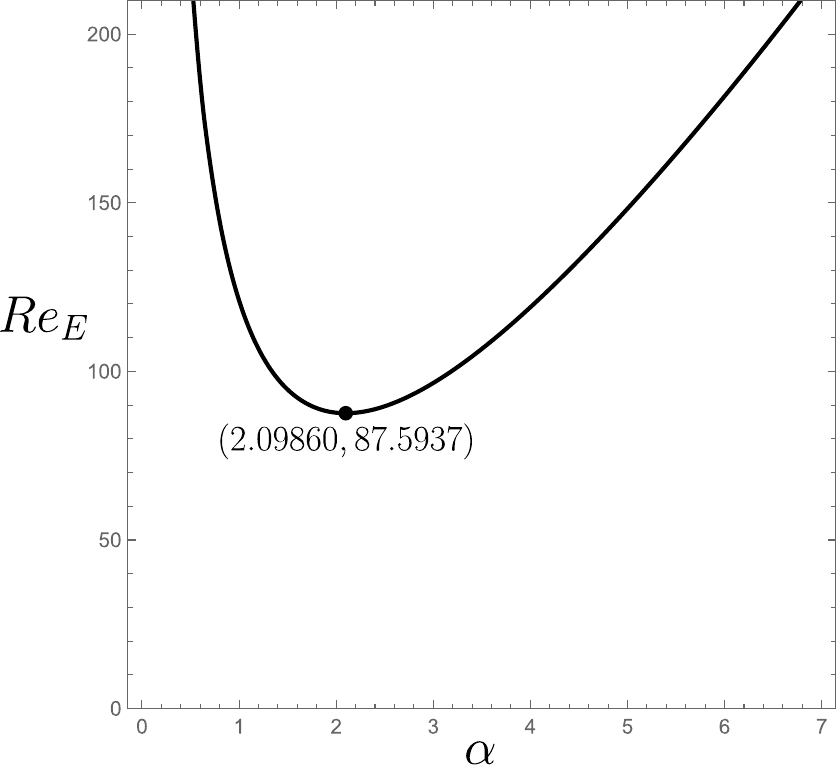}
\caption{\label{fig1}Poiseuille flow and T-modes: Solution of \eqref{47} in the plane $\qty(\alpha, Re_E)$}
\end{figure}

\begin{figure}[t]
\centering
\includegraphics[width=0.5\textwidth]{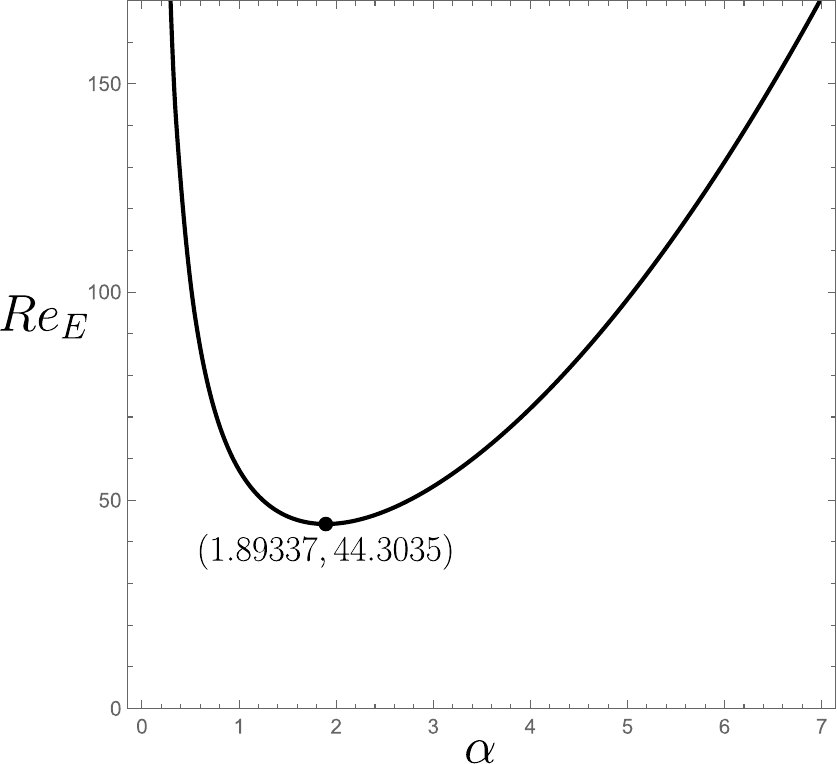}
\caption{\label{fig2}Couette flow and T-modes: Solution of \eqref{47} in the plane $\qty(\alpha, Re_E)$}
\end{figure}

On account of \eqref{40} and \eqref{42}, the partial differential equation \eqref{44} is subjected to the boundary conditions
\eq{45}{
z = \pm 1 : \qquad \pdv{\psi}{x} = 0 \qc \pdv{\psi}{z} = 0 .
}

\subsection{Normal modes}
We introduce a normal mode expression for the solution of \eqref{44} and \eqref{45},
\eq{46}{
\psi(x,z) = \Psi(z) \, e^{i\, \alpha x} ,
}
where $\alpha = \alpha_x$ is the wavenumber and the time dependence has been ignored as \eqref{44} and \eqref{45} are time independent. By substituting \eqref{46} in \eqref{44} and \eqref{45} and by employing \eqref{27}, we obtain
\eq{47}{
\begin{cases}
\displaystyle
\Psi'''' - 2 \alpha^2 \Psi'' + \alpha^4 \Psi + i \, \alpha \, Re_E \qty(f' \Psi' + \frac{1}{2} f'' \Psi ) = 0,\\[6pt]
\Psi(\pm 1) = 0 \qc \Psi'(\pm 1) = 0 .
\end{cases}
}
The T-modes eigenvalue problem \eqref{47} can be solved by a series expansion of $\Psi(z)$ based on  Chebyshev polynomials and Galerkin's weighted residuals method, following the steps described in \citet{barletta2024alternative} and reported in Appendix \ref{AppA}. In fact, we employ the expansion
\eq{48}{
\Psi(z) = \qty(1 - z^2)^2 \,\sum_{n=0}^\infty q_n \,T_{n}(z) ,
}
where $q_n$ are constant complex-valued coefficients that uniquely determine the eigenfunction $\Psi(z)$.
The results of the numerical solution are reported graphically in Figs.~\ref{fig1} and \ref{fig2} for Poiseuille flow and Couette flow, respectively. 

\begin{table}[t]
\centering
\begin{tabular}{|c|c|}
\hline
Poiseuille flow & Couette flow\\
\hline\hline
~~~~~~$\pm 87.7435$ & ~~~~~~$\pm 44.4126$ \\
\hline
~~~$\pm 168.486$ & ~~~$\pm 188.554$ \\
\hline
~~~$\pm 802.943$ & ~~~$\pm 534.833$ \\
\hline
$\pm 1179.97$ & $\pm 1176.01$ \\
\hline
$\pm 3026.08$ & $\pm 2204.94$ \\
\hline
$\pm 3918.81$ & $\pm 3714.59$ \\
\hline
$\pm 7638.66$ & $\pm 5797.92$ \\
\hline
\end{tabular}
\caption{\label{tab1}\centering{}T-modes eigenvalue problem with $\alpha=2$: lowest seven\linebreak eigenvalues $Re_E$, ordered by increasing absolute value}
\end{table}

The effective upper bound to nonlinear stability is defined by the minima of the curves displayed in the $\qty(\alpha, Re_E)$ plane. Hence, the T-modes eigenvalue problem yields $Re_E = 87.5937$ for Poiseuille flow and $Re_E = 44.3035$ for Couette flow. Such minima are achieved with $\alpha = 2.09860$ and $\alpha =1.89337$ for Poiseuille and Couette flows, respectively.

\subsection{Further remarks}\label{remarkOrr}
Equation \eqref{39} entails $\hat{v}(x,z)$ being a solution of the Laplace equation subject to the boundary conditions \eqref{40}, together with periodicity over $x \in \qty[0, 2\pi/\alpha]$. By a Fourier series expansion in $x$, it can be proved that the only possible case is $\hat{v}(x,z)=0$, meaning that the velocity field for the T-modes eigenvalue problem is two-dimensional with just two nonzero $y$-independent components. 

An interesting property of the eigenvalue problem defined by \eqref{47} is obtained by taking its complex conjugate, 
\eq{47cc}{
\begin{cases}
\displaystyle
\bar\Psi'''' - 2 \alpha^2 \bar\Psi'' + \alpha^4 \bar\Psi - i \, \alpha \, Re_E \qty(f' \bar\Psi' + \frac{1}{2} f'' \bar\Psi ) = 0,\\[6pt]
\bar\Psi(\pm 1) = 0 \qc \bar\Psi'(\pm 1) = 0 ,
\end{cases}
}
where $\bar\Psi(z)$ is the complex conjugate of $\Psi(z)$. Hence, one may justify the mathematical reason behind the eigenvalue pairs $Re_E$, reported in Table~\ref{tab1}, having the same absolute values but opposite signs. In fact, for every given $\alpha$, \eqref{47} and \eqref{47cc} yield eigenvalues $Re_E$ and their opposites $-Re_E$ with eigenfunctions given by $\Psi(z)$ and $\bar\Psi(z)$, respectively.

\section{L-modes eigenvalue problem}\label{joseph}
Let us now assume that both $\vb{\hat{u}}$ and $\lambda$ are independent of $x$. Hereafter, such a case will be termed {\em longitudinal modes} and abbreviated with L-modes. Then, the zero-divergence condition on $\vb{\hat{u}}$ can be expressed as,
\eq{49}{
\pdv{\hat{v}}{y} + \pdv{\hat{w}}{z} = 0,
}
which is identically satisfied on defining a streamfunction $\psi$ such that
\eq{50}{
\hat{v} = \pdv{\psi}{z} \qc \hat{w} = - \pdv{\psi}{y} .
}
By employing \eqref{50}, \eqref{39} yield
\eq{51}{
\begin{cases}
\displaystyle
2 m\, \laplacian \hat{u} + \pdv{\psi}{y}\, f' = 0, \\[12pt] 
\displaystyle
2 m\, \laplacian{\pdv{\psi}{z}} = \pdv{\lambda}{y},\\[12pt]
\displaystyle
2m\, \laplacian{\pdv{\psi}{y}} + \hat{u} \, f' = - \pdv{\lambda}{z} .
\end{cases}
}
By deriving the second \eqref{51} with respect to $z$, the third \eqref{51} with respect to $y$ and summing the two resulting equations, we encompass the dependence on the Lagrange multiplier $\lambda$ and we get 
\eq{52}{
\begin{cases}
\displaystyle
2 m\, \laplacian \hat{u} + \pdv{\psi}{y}\, f' = 0, \\[12pt] 
\displaystyle
2 m\, \nabla^4{\psi} + \pdv{\hat{u}}{y}\, f' = 0 .
\end{cases}
}
On account of \eqref{40} and \eqref{50}, the system of partial differential equations \eqref{52} is subjected to the boundary conditions
\eq{53}{
z = \pm 1 : \qquad \hat{u} = 0 \qc \pdv{\psi}{y} = 0 \qc \pdv{\psi}{z} = 0 .
}

\begin{figure}[t]
\centering
\includegraphics[width=0.5\textwidth]{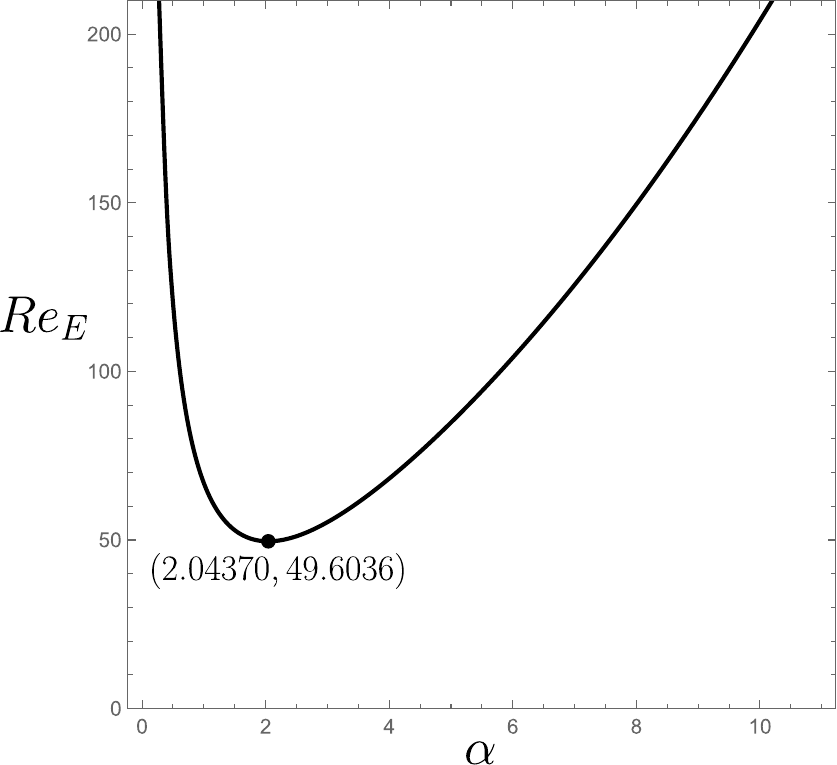}
\caption{\label{fig3}Poiseuille flow and L-modes: Solution of \eqref{55} in the plane $\qty(\alpha, Re_E)$}
\end{figure}

\begin{figure}[t]
\centering
\includegraphics[width=0.5\textwidth]{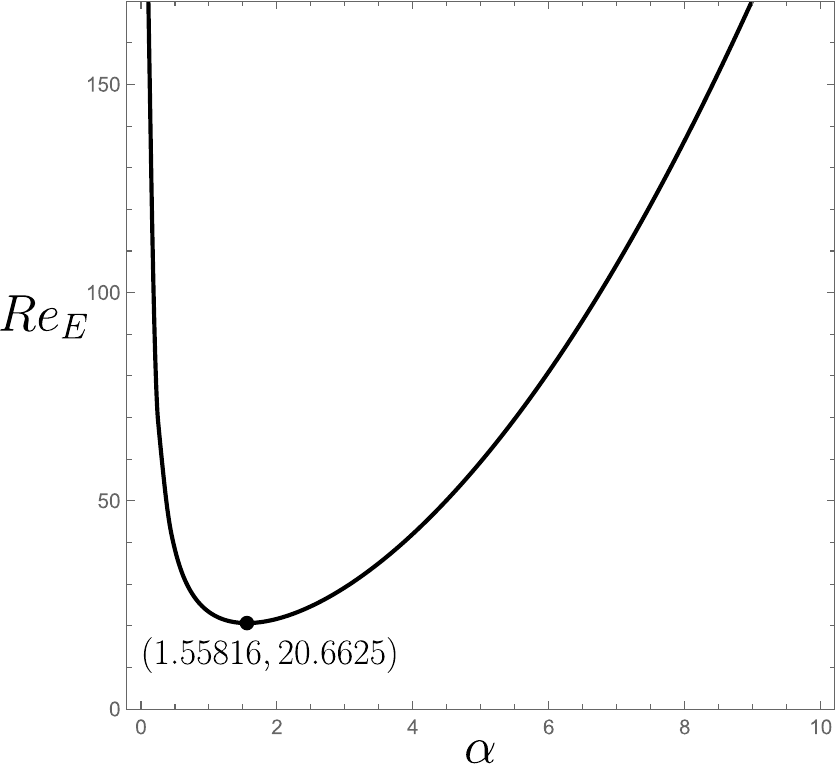}
\caption{\label{fig4}Couette flow and L-modes: Solution of \eqref{55} in the plane $\qty(\alpha, Re_E)$}
\end{figure}

\subsection{Normal modes}
We introduce a normal mode expression for the solution of \eqref{52} and \eqref{53},
\eq{54}{
\hat{u}(y,z) = \eta(z) \, e^{i\, \alpha y} \qc \psi(y,z) = \Psi(z) \, e^{i\, \alpha y} ,
}
where $\alpha = \alpha_y$ is the wavenumber which, in this case, serves to model the periodicity along the $y$ axis. Again, the time dependence has been ignored as \eqref{52} and \eqref{53} define a stationary problem. By substituting \eqref{54} in \eqref{52} and \eqref{53} and by employing \eqref{27}, we obtain
\eq{55}{
\begin{cases}
\displaystyle
2 \, \qty(\eta'' - \alpha^2\, \eta) + i\, \alpha\,Re_E\, \Psi\, f' = 0, \\[6pt] 
\displaystyle
2 \, \qty(\Psi'''' - 2 \alpha^2\, \Psi'' + \alpha^4\, \Psi) + i\, \alpha\,Re_E\, \eta \, f' = 0 ,\\[6pt]
\eta(\pm 1) = 0 \qc \Psi(\pm 1) = 0 \qc \Psi'(\pm 1) = 0 .
\end{cases}
}
The L-modes eigenvalue problem \eqref{47} can be solved by employing series expansions of both $\eta(z)$ and $\Psi(z)$ based on  Chebyshev polynomials and Galerkin's weighted residuals method, according to
\eq{56}{
\eta(z) = \qty(1 - z^2) \,\sum_{n=0}^\infty s_n \,T_{n}(z) , \nn\\
\Psi(z) = \qty(1 - z^2)^2 \,\sum_{n=0}^\infty q_n \,T_{n}(z) ,
}
where $s_n$ and $q_n$ are complex-valued series coefficients determined by \eqref{55}. As a consequence of \eqref{56}, both $\eta(z)$ and $\Psi(z)$ satisfy the boundary conditions specified in \eqref{55}.

\begin{table}[t]
\centering
\begin{tabular}{|c|c|}
\hline
Poiseuille flow & Couette flow\\
\hline\hline
~~~~~~$\pm 49.6203$ & ~~~~~~$\pm 21.6752$ \\
\hline
~~~~~~$\pm 61.7493$ & ~~~~~~$\pm 70.1509$ \\
\hline
~~~$\pm 285.569$ & ~~~$\pm 173.451$ \\
\hline
~~~$\pm 342.711$ & ~~~$\pm 354.460$ \\
\hline
~~~$\pm 931.242$ & ~~~$\pm 636.330$ \\
\hline
$\pm 1057.86$ & $\pm 1042.27$ \\
\hline
$\pm 2207.40$ & $\pm 1595.52$ \\
\hline
\end{tabular}
\caption{\label{tab2}\centering{}L-modes eigenvalue problem with $\alpha=2$: lowest seven\linebreak eigenvalues $Re_E$, ordered by increasing absolute value}
\end{table}

The results of the numerical solution are reported graphically in Figs.~\ref{fig3} and \ref{fig4} for Poiseuille flow and Couette flow, respectively. For every assigned $\alpha$, one has a sequence of eigenvalues where eigenvalue pairs exist sharing the same absolute value with both the plus and minus signs. Table~\ref{tab2} reports such a sequence ordered by increasing absolute values and limited to the lowest seven pairs. The threshold curves in the $\qty(\alpha, Re_E)$ plane shown in  Figs.~\ref{fig3} and \ref{fig4} are obtained by taking for each $\alpha$ the smallest positive eigenvalue $Re_E$. 

The effective upper bound to nonlinear stability is defined by the minima of the curves displayed in the $\qty(\alpha, Re_E)$ plane (see Figs.~\ref{fig3} and \ref{fig4}). Hence, the L-modes eigenvalue problem yields $Re_E = 49.6036$ for Poiseuille flow and $Re_E = 20.6625$ for Couette flow. Such minima are achieved with $\alpha = 2.04370$ and $\alpha = 1.55816$ for Poiseuille and Couette flows, respectively. The results for Poiseuille flow agree completely with the data reported by \citet{xiong2020linear}.

\subsection{Further remarks}
\subsubsection{Three-dimensional velocity field}
The eigenfunctions of problem \eqref{55} are such that $\eta(z)$ and, hence, also $\hat{u}(y,z)$ do not vanish identically. In fact, if $\eta(z)$ were zero, then the first \eqref{55} would imply that $\Psi(z)$ is also identically zero, which means the trivial solution of \eqref{55}. Therefore, the solution of the eigenvalue problem involves a three-dimensional velocity field $\vb{\hat{u}}$ which, however, depends on just two coordinates, namely $(y,z)$.

\subsubsection{The Rayleigh-B\'enard analogy}
If one considers Coutte flow, $f'(z) = 1$, by defining
\eq{58}{
\Phi = 2\, i\,\alpha\,Re_E\, \Psi \qc \xi = \frac{z+1}{2} \qc \beta = 2\alpha,
}
then \eqref{55} can be rewritten as
\eq{59}{
\begin{cases}
\displaystyle
\dv[2]{\eta}{\xi} - \beta^2\, \eta + \Phi = 0, \\[12pt] 
\displaystyle
\dv[4]{\Phi}{\xi} - 2 \beta^2\, \dv[2]{\Phi}{\xi} + \beta^4\, \Phi - 4\,\beta^2\, Re_E^2\, \eta = 0 ,\\[6pt]
\displaystyle
\xi=0,1:\qquad \eta = 0 \qc \Phi = 0 \qc \dv{\Phi}{\xi} = 0 .
\end{cases}
}
In the form \eqref{59}, L-modes eigenvalue problem coincides with the Rayleigh-B\'enard eigenvalue problem for the linear instability in a horizontal fluid layer with heating from below and subjected to rigid-wall no-slip conditions at the plane isothermal boundaries (see, for example, Section~7.5 of \citet{barletta2019routes}). The coincidence is perfect if we define 
\eq{60}{
Ra = 4 \, Re_E^2 ,
}
instead of the Rayleigh number. As is well-known, the numerical solution for the linear instability analysis relative the Rayleigh-B\'enard problem yields the critical values \cite{barletta2019routes}
\eq{61}{
\beta_c = 3.11632 \qc Ra_c = 1707.76 ,
}
corresponding to the minimum of the neutral stability curve in the $(\beta, Ra)$ plane. By using these numerical data, on account of \eqref{58} and \eqref{60}, one has
\eq{62}{
\alpha = \frac{3.11632}{2} = 1.55816 \qc Re_E =\frac{1}{2} \sqrt{1707.76} = 20.6625 ,
}
in perfect agreement with the data for the minimum of the curve in the $\qty(\alpha, Re_E)$ plane reported in Fig.~\ref{fig4}. Incidentally, \eqref{59} and \eqref{60} offer a pretty simple justification for the eigenvalues found for problem \eqref{55}. In fact, when such a problem is rewritten in the form \eqref{59}, the eigenvalue $Re_E$ is present only through its square so that both $\pm|Re_E|$ are allowed. This reasoning holds for the Couette flow, but one may write an eigenvalue problem analogous to \eqref{59} for the Poiseuille flow,
\eq{63}{
\begin{cases}
\displaystyle
\dv[2]{\eta}{\xi} - \beta^2\, \eta - 2\, \qty(2\, \xi - 1) \Phi = 0, \\[12pt] 
\displaystyle
\dv[4]{\Phi}{\xi} - 2 \beta^2\, \dv[2]{\Phi}{\xi} + \beta^4\, \Phi + 8\, \qty(2\, \xi - 1)\,\beta^2\, Re_E^2\, \eta = 0 ,\\[6pt]
\displaystyle
\xi=0,1:\qquad \eta = 0 \qc \Phi = 0 \qc \dv{\Phi}{\xi} = 0 ,
\end{cases}
}
so that the argument to justify the eigenvalue pairs $\pm|Re_E|$ can be promptly extended also to the case of Poiseuille flow.

\begin{figure}[t]
\centering
\includegraphics[width=0.8\textwidth]{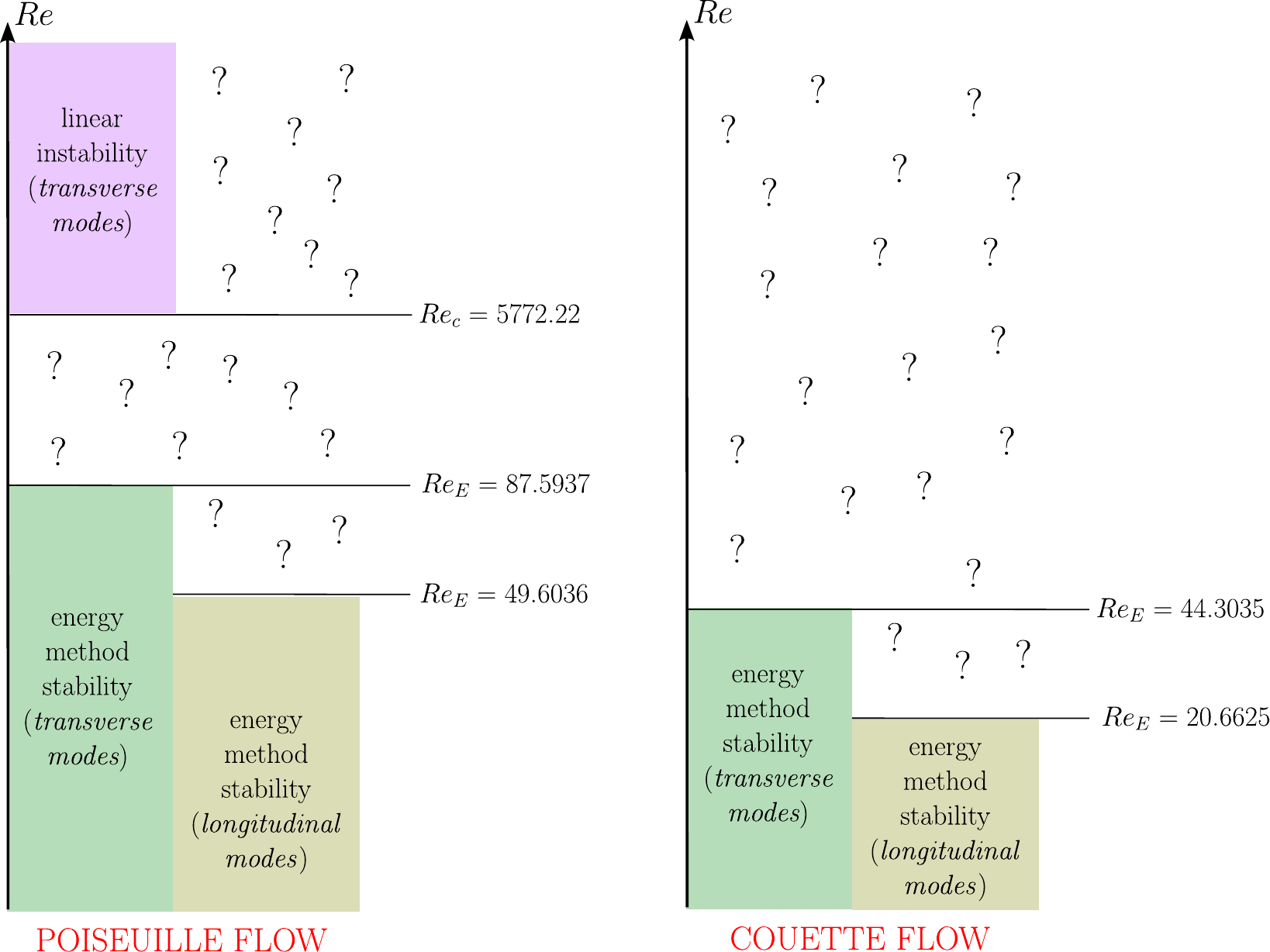}
\caption{\label{fig5}\centering{}A sketch of the stability and instability information gathered\linebreak from the linear analysis and from the energy method analysis}
\end{figure}

\section{Discussion of the results}\label{dire}
The first obvious argument to be pointed out is that the energy method defines parametric thresholds in terms of the Reynolds number completely different from the critical values obtained through the linear instability analysis \cite{orszag1971accurate, zebib1984chebyshev, barletta2024alternative}. A schematic description of the stability and instability information gathered from the linear analysis and from the energy method analysis about plane Poiseuille and Couette flows is provided in Fig.~\ref{fig5}. As is well-known \cite{orszag1971accurate, zebib1984chebyshev, barletta2024alternative}, the linear analysis provides the critical Reynolds number, $Re_c$, for the onset of the linear instability relative to transverse modes perturbing the Poiseuille flow. On the other hand, no such critical values exist either for Poiseuille flow with longitudinal modes or for Couette flow whatever might be the type of linear perturbing modes \cite{romanov1973stability}. The cloud of question marks in Fig.~\ref{fig5} fills the ranges of Reynolds number left undetermined by either the linear method or the energy method. In fact, for such intermediate Reynolds numbers, one may well have finite amplitude (nonlinear) instability which, however, cannot be detected with the linear analysis or the energy method. There is evidence documenting such a nonlinear instability coming, for instance, from experiments \cite{nishioka1975experimental, carlson1982flow, tillmark1992experiments, daviaud1992subcritical, prigent2003long, barkley2007mean}.

The correct interpretations of the linear stability analysis and of the energy method analysis are that the former defines a lower bound to the parametric region where a stable evolution of perturbations is impossible, while the latter defines an upper bound to the parametric region where an unstable evolution of perturbations is impossible. What happens in between those parametric bounds is merely undetermined from both the linear method and the energy method standpoints.

A much subtler aspect of the energy method regards the role played by the Euler-Lagrange equations \eqref{39}. They were obtained as a consequence of a maximum problem formulation where the qualifying point is the specification of the set $\cal S$ where the functional maximum $m$ defined by \eqref{21} is sought. The elements of $\cal S$ have been called the kinematically admissible velocity fields $\vb{u}=(u, v, w)$. The term kinematically admissible means two requirements that must be satisfied by $\vb{u}$, the fulfilment of the boundary conditions and the condition of a zero divergence. The solution of the Euler-Lagrange equations, $\vb{\hat{u}} = \qty(\hat{u}, \hat{v}, \hat{w})$ obviously must be an element of $\cal S$, but should $\vb{\hat{u}}$ also be a solution of the local momentum balance equations given by \eqref{6} with a suitable pressure field $\hat{p}$\,? The general answer is no, meaning that a kinematically admissible velocity field, in general, is not a dynamically admissible velocity field satisfying all equations \eqref{6}. The proof can be found with {\em reductio ad absurdum}.
In fact, let $\vb{\hat{u}} = \qty(\hat{u}, \hat{v}, \hat{w})$ be a solution of \eqref{39} for $m = 1/Re_E$ and let us assume that $\vb{\hat{u}} = \qty(\hat{u}, \hat{v}, \hat{w})$ be dynamically admissible so that it satisfies the local momentum balance given by \eqref{6} with $Re = Re_E$. Due to the argument presented in Section~\ref{propm}, we can well neglect the nonlinear terms in \eqref{6} as their contribution does not impact on the evaluation of $m$ and, hence, on the Euler-Lagrange equations and their solution. Furthermore, on carrying out the substitution of $\vb{\hat{u}} = \qty(\hat{u}, \hat{v}, \hat{w})$ into \eqref{6}, the time-dependence turns out to be important, whereas it had been immaterial on dealing with the Euler-Lagrange equations \eqref{39}. Thus, we introduce a factor $e^{\gamma t}$, with $\gamma \in \mathbb{C}$, multiplying the otherwise stationary solution $\vb{\hat{u}} = \qty(\hat{u}, \hat{v}, \hat{w})$. With such specifications, the substitution of $\vb{\hat{u}}\, e^{\gamma t}$ and $\hat{p}\, e^{\gamma t}$, where $\hat{p}$ is a stationary pressure field, into \eqref{6} leads us to the following system of momentum balance equations
\eq{67}{
\begin{cases}
\displaystyle
\gamma\, \hat{u} + f \pdv{\hat{u}}{x} + \hat{w} f' = - \pdv{\hat{p}}{x} + \frac{1}{Re_E} \laplacian{\hat{u}} ,\\[12pt]
\displaystyle
\gamma\, \hat{v} + f \pdv{\hat{v}}{x}  = - \pdv{\hat{p}}{y} + \frac{1}{Re_E} \laplacian{\hat{v}} , \\[12pt]
\displaystyle
\gamma\, \hat{w} + f \pdv{\hat{w}}{x} = - \pdv{\hat{p}}{z} + \frac{1}{Re_E} \laplacian{\hat{w}} . 
\end{cases} 
}
Since $\vb{\hat{u}}$ is a solution of \eqref{39}, we can employ \eqref{39} to express $\laplacian{\vb{\hat{u}}}$ and substitute it into \eqref{67},
\eq{68}{
\begin{cases}
\displaystyle
\gamma\, \hat{u} + f \pdv{\hat{u}}{x} + \frac{1}{2}\, \hat{w} f' = - \pdv{\Lambda}{x}  ,\\[12pt]
\displaystyle
\gamma\, \hat{v} + f \pdv{\hat{v}}{x}  = - \pdv{\Lambda}{y}  , \\[12pt]
\displaystyle
\gamma\, \hat{w} + f \pdv{\hat{w}}{x} -  \frac{1}{2}\, \hat{u} f' = - \pdv{\Lambda}{z} , 
\end{cases} 
}
where
\eq{69}{
\Lambda = \hat{p} - \frac{1}{2}\, \lambda .
}
We now multiply the first \eqref{68} by the complex conjugate of $\hat{u}$, the second \eqref{68} by the complex conjugate of $\hat{v}$ and the third \eqref{68} by the complex conjugate of $\hat{w}$, we integrate over $\Omega$ and we sum the three resulting equations. The result is
\eq{70}{
\gamma \, \norm{\vb{\hat{u}}}^2 + i\, \alpha_x \braket{f}{|\vb{\hat{u}}|^2} + i\, \Im\!\qty(\braket{\hat{u}}{\hat{w}\, f'}) = 0 ,
}
where $\Im(\cdot)$ denotes the imaginary part of a complex quantity.
On writing \eqref{70}, the scalar product $\braket{\cdot\,}{\cdot}$ is meant to be defined as 
\eq{65}{
\bra{g} \ket{h}  = \int\limits_{\Omega} \bar{g}\, h\ \dd^3 x,
}
and $|\vb{\hat{u}}|^2$ is the square modulus of a complex-valued vector field. Furthermore, \eqref{70} relies on a modal form of the eigenfunction $\vb{\hat{u}}$ where the dependence on $x$ and $y$ can be factored out with an exponential term, $\exp\!\qty(i \,\alpha_x\, x + i\, \alpha_y\, y)$. Finally, the term
\[
\braket{\hat{u}_j}{\pdv{\Lambda}{x_j}}
\]
can be proved to be zero with a reasoning similar to that reported in \eqref{15}. In \eqref{70}, all the quantities $\norm{\vb{\hat{u}}}^2$, $\braket{f}{|\vb{\hat{u}}|^2}$ and $\Im\!\qty(\braket{\hat{u}}{\hat{w}\, f'})$ are evidently real. Thus, one can deduce that the real part of the complex growth rate $\gamma$ must be zero so that $\vb{\hat{u}}\, e^{\gamma\, t}$ does not decay in time as predicted by the linear stability analysis for the subcritical regime \cite{orszag1971accurate, zebib1984chebyshev, barletta2024alternative}. Indeed, we have found that $Re_E$ is smaller than $Re_c$ both for Poiseuille flow and for Couette flow. Hence, our conclusion is that forcing $\vb{\hat{u}}\, e^{\gamma\, t}$ to be dynamically admissible leads to a contradiction. 

The proof reported above applies to plane shear flows, but one can devise a general idea that the Euler-Lagrange equations obtained for the determination of the energy method threshold for stability are generally incompatible with the linearised perturbation equations employed for the determination of the linear threshold to instability, or neutral stability condition. There is an obvious exception to the rule: those cases where the two thresholds coincide, meaning that the critical value to instability coincides with the energy method threshold to stability. Such exceptions are quite important as they address situations where no subcritical instability is possible. A well-known example is the Rayleigh-B\'enard problem, as shown for instance in \citet{joseph1976stability2} or in \citet{straughan2013energy}.

Another observation is that the classical results reported above, {\em i.e.} the critical nonlinear
 Reynolds numbers $Re_E$ obtained on L-modes and on T-modes do not solve the maximum problem in the general case for 3-dimensional perturbations (three components of the velocity field dependent on the three variables $x,y,z$), but are limited only to the case of dependence on two spatial variables, the limiting cases $2.5$-dimensional, $u(y,z), v(y,z) , w(y,z)$, and $2$-dimensional, $u(x,z), 0, w(x,z)$.
To be sure that the critical Reynolds energy value obtained is the optimal one, it is necessary to first test on which perturbations the maximum is obtained. Recent works by Mulone \cite{Mulone2024}, Falsaperla, Mulone and Perrone \cite{FALSAPERLA202293} prove that this maximum is obtained on the 2-dimensional transverse perturbations (see Orr \cite{Orr1907}). For Reynolds numbers lower than this critical value all 3-dimensional, 2.5-dimensional and 2-dimensional perturbations have an energy that decreases monotonically with time. This leads to the validity also for the nonlinear system of a Squire theorem.
Furthermore, for the longitudinal case in Falsaperla, Mulone and Perrone \cite{FALSAPERLA202293} and in 
Mulone \cite{Mulone2024}, it is proven that the longitudinal dynamic perturbations are always stabilizing (as in the linear case).
Classical results giving sufficient conditions of nonlinear stability for critical Reynolds numbers lower than the critical value found for L-modes are compatible with these recent results.
This interesting problem still requires careful examination and in-depth analysis, also with a view at finding the critical perturbations that can give values close to those of the numerical analysis and experiments \cite{PhysRevE.100.013113}.

\section{Conclusions}
A pedagogical introduction to the energy method and its exploitation for the nonlinear stability analysis of the Couette and Poiseuille flows in a plane-parallel channel have been provided. The details in the determination of the time evolution undergone the energy functional and the fundamental inequality satisfied by the time derivative of the energy have been presented step-by-step. The optimisation of such inequality as achieved by functional methods leading to an extremum condition and, thus, to the Euler-Lagrange equations has been discussed. Finally, the two cases of transverse modes and longitudinal modes have been treated, showing up all the details of the numerical solutions yielding the energy bound to the stability region. This tutorial is intended to provide all the needed information to understand the topic, together with the numerical tool and software code necessary to gather the relevant numerical data. The results obtained in terms of threshold conditions in the parametric $(\alpha, Re_E)$ plane have been critically examined in order to disclose the subtleties and possible misunderstandings regarding their physical meaning. The main message that, sometimes, is not so neatly outlined is that the linear threshold to instability, also known as neutral stability condition, and the energy method threshold to stability are insufficient to support any statement regarding the existence of subcritical instability. Then, any statement about the onset of a subcritical instability for either the Couette or the Poiseuille flows is a pure conjecture. Making such a conjecture a deduction means going beyond the modal linear instability analysis and the nonlinear stability analysis carried out via the energy method. One should rely on the transient growth analysis of non-modal perturbations or on the weakly nonlinear analysis of the instability or, possibly, on the direct numerical simulation of the nonlinear time-evolution of the flow. In particular, the diverse energy method thresholds to stability achieved by considering transverse modes or longitudinal modes do not imply a priori that the onset of a possible subcritical instability happens necessarily through those modes whose energy method threshold is the lowest.

\subsection*{Acknowledgements}
The work by Antonio Barletta was supported by the University of Bologna, grant number RFO-2023.

\subsection*{Authors' contributions}
A. Barletta, G. Mulone: these authors contributed equally to this work.

\bibliographystyle{elsarticle-num-names}
\bibliography{biblio}

\begin{appendices}

\section{Solution of T-modes eigenvalue problem}\label{AppA}
\setcounter{equation}{0}
\renewcommand{\theequation}{\thesection.\arabic{equation}}
\setcounter{table}{0}
\renewcommand{\thetable}{\thesection.\arabic{table}}

In order to solve \eqref{47}, we introduce test functions that identically satisfy the boundary conditions at $z=\pm1$,
\eq{A1}{
\varphi_n(z) = (1-z^2)^2\, T_n(z)\qc n=0,1,2,3,\ \ldots\ ,
}
where $T_n(z)$ is the Chebyshev polynomial of order $n$. We expand $\Psi(z)$ as shown in \eqref{48}. Then, the ordinary differential equation \eqref{47} can be rewritten approximately as a sum over the first $N+1$ terms, where $N$ is the truncation order,
\eq{A2}{
\sum_{n=0}^N q_n \qty[\varphi''''_n - 2 \alpha^2 \varphi''_n + \alpha^4 \varphi_n + i \, \alpha \, Re_E \qty(f' \varphi'_n + \frac{1}{2} f'' \varphi_n )] = 0.
}
We now multiply \eqref{A2} by $\varphi_m(z)$ and then we integrate over $z \in [-1,1]$, so that we obtain
\eq{A3}{
\sum_{n=0}^N {\cal A}_{m n}\, q_n  = Re_E \sum_{n=0}^N {\cal B}_{m n}\, q_n \qc \text{namely} \quad {\boldsymbol{\cal A}} \vdot \vb{q} = Re_E\, {\boldsymbol{\cal B}}\vdot \vb{q} ,
}
where symbols ${\boldsymbol{\cal A}}$, ${\boldsymbol{\cal B}}$ and $\vb{q}$ denote $N+1$ dimensional square matrices and array. Within this appendix, symbol $m$ is meant to denote a non-negative  integer just as $n$, as there is no possibility to confuse it with the same symbol $m$ defined by \eqref{21} and not relevant for this appendix. The matrix elements of ${\boldsymbol{\cal A}}$ and ${\boldsymbol{\cal B}}$ are given by 
\eq{A4}{
{\cal A}_{m n} = \braket{\varphi_m}{\varphi''''_n} - 2 \alpha^2 \braket{\varphi_m}{\varphi''_n} + \alpha^4 \braket{\varphi_m}{\varphi_n}  , \nn\\
\nn\\
{\cal B}_{m n} = - i \, \alpha  \qty(\braket{\varphi_m}{f' \varphi'_n} + \frac{1}{2} \braket{\varphi_m}{f'' \varphi_n }).
} 
Here, the scalar product $\braket{\cdot\,}{\cdot}$ is just meant to be an integration over $z \in [-1,1]$.
The generalised eigenvalue problem defined by \eqref{A3}, where $Re_E$ is the eigenvalue, can be solved by employing the software {\sl Mathematica 14} ({\small \copyright}\ Wolfram Research, Inc.). We refer to the Poiseuille flow, as the case of Couette flow differs just in the expression of $f(z)$. The first step is defining the matrix elements of ${\boldsymbol{\cal A}}$ and ${\boldsymbol{\cal B}}$ after fixing a maximum value for $N$, say $N=10$,

\begin{table}[t]
\centering
\begin{tabular}{|c|c|c|}
\hline
number of terms & $Re_E$ & $Re_E$\\
\hline\hline
$8+1$   & 87.74360857601487 & 168.5289905466895\\
$10+1$ & 87.74349216899427 & 168.4871619060524\\
$12+1$ & 87.74349116538390 & 168.4859740388069\\
$14+1$ & 87.74349115949544 & 168.4859578335054\\
$16+1$ & 87.74349115946921 & 168.4859576616534\\
$18+1$ & 87.74349115946914 & 168.4859576604801\\
$20+1$ & 87.74349115946914 & 168.4859576604742\\
$22+1$ & 87.74349115946914 & 168.4859576604742\\
\hline
\end{tabular}
\caption{\label{tabA1}\centering{}T-modes eigenvalue problem for the Poiseuille flow, with $\alpha=2$: \linebreak the smallest two positive eigenvalues $Re_E$, with increasing \linebreak number of terms in the series approximation}
\end{table}

%
%

\begin{footnotesize}
\begin{verbatim}
NN=10; f[z_]:=1-z^2; phi[n_,z_]:=(1-z^2)^2 ChebyshevT[n,z];
Do[A0[m,n]=Integrate[phi[m,z] phi[n,z],{z,-1,1}],{m,0,NN},{n,0,NN}];
Do[A2[m,n]=Integrate[phi[m,z] D[phi[n,z],{z,2}],{z,-1,1}],{m,0,NN},{n,0,NN}];
Do[A4[m,n]=Integrate[phi[m,z] D[phi[n,z],{z,4}],{z,-1,1}],{m,0,NN},{n,0,NN}];
Do[AU1[m,n]=Integrate[phi[m,z] f'[z] D[phi[n,z],z],{z,-1,1}],{m,0,NN},{n,0,NN}];
Do[AU2[m,n]=Integrate[phi[m,z] f''[z] phi[n,z],{z,-1,1}],{m,0,NN},{n,0,NN}];
AA[alpha_,M_]:=Table[A4[m,n]-2 alpha^2 A2[m,n]+alpha^4 A0[m,n],{m,0,M},{n,0,M}];
BB[alpha_,M_]:=Table[-(I alpha AU1[m,n]+1/2 I alpha AU2[m,n]),{m,0,M},{n,0,M}];
\end{verbatim}
\end{footnotesize}

\noindent{}If the value of $N$ is fixed, one can always test the accuracy of the approximation for every truncation order less or equal than $N$. For instance, one may obtain the eigenvalues for $\alpha = 2$ and a truncation to the first $8+1$ terms,

\begin{footnotesize}
\begin{verbatim}
NumberForm[SortBy[With[{alpha=N[2,40],M=8},Chop[Eigenvalues[{AA[alpha,M],BB[alpha,M]}]]],
N[Abs[Re[#]]]&],6]
\end{verbatim}
\end{footnotesize}

\noindent{}The output for $Re_E$ is

\begin{footnotesize}
\begin{verbatim}
{-87.7436,87.7436,-168.529,168.529,-836.388,836.388,-1581.69,1581.69,Infinity}
\end{verbatim}
\end{footnotesize}

\noindent{}The eigenvalue $Re_E = \infty$ may just be ignored. By comparing these data with Table~\ref{tab1}, one immediately sees that with such a few terms, the approximation of the smallest eigenvalue pair $\pm|Re_E|$ is good, while the accuracy for eigenvalues with a larger absolute value rapidly decreases. A test of the convergence in the evaluation of the smallest two positive eigenvalues is reported in Table~\ref{tabA1}.

\section{Solution of L-modes eigenvalue problem}\label{AppB}
\setcounter{equation}{0}
\setcounter{table}{0}

In order to solve \eqref{55}, we introduce test functions that identically satisfy the boundary conditions at $z=\pm1$,
\eq{B1}{
\varphi_n(z) = (1-z^2)^2\, T_n(z)\qc \tilde\varphi_n(z) = (1-z^2)\, T_n(z)\qc \quad  n=0,1,2,3,\ \ldots\ .
}
We now expand $\eta(z)$ and $\Psi(z)$ according to \eqref{56}. Then, the system of two ordinary differential equations \eqref{55} can be rewritten approximately as a sum over the first $N+1$ terms, where $N$ is the truncation order,
\eq{B2}{
\begin{cases}
\displaystyle
\sum_{n=0}^N s_n\, \qty(\tilde\varphi''_n - \alpha^2\, \tilde\varphi_n) + i\, \frac{\alpha}{2}\,Re_E\, f' \, \sum_{n=0}^N q_n\, \varphi_n = 0, \\[12pt] 
\displaystyle
\sum_{n=0}^N q_n\, \qty(\varphi''''_n - 2 \alpha^2\, \varphi''_n + \alpha^4\, \varphi_n) + i\, \frac{\alpha}{2}\,Re_E\,  \, f' \sum_{n=0}^N s_n\, \tilde\varphi_n = 0 .
\end{cases}
}
We multiply the first \eqref{B2} by $\tilde\varphi_m(z)$ and then we integrate over $z \in [-1,1]$. We multiply the second \eqref{B2} by $\varphi_m(z)$ and then we integrate over $z \in [-1,1]$. 
Within this appendix, symbol $m$ is meant to denote a non-negative  integer, as there is no possibility to confuse it with the same symbol $m$ defined by \eqref{21} and not relevant for this appendix. 
Hence, we can write
\eq{B3}{
\begin{cases}
\displaystyle
\sum_{n=0}^N s_n \qty(\braket{\tilde\varphi_m}{\tilde\varphi''_n} - \alpha^2 \braket{\tilde\varphi_m}{\tilde\varphi_n}) + i\, \frac{\alpha}{2}\,Re_E\, \sum_{n=0}^N q_n \braket{\tilde\varphi_m}{f'\, \varphi_n} = 0, \\[12pt] 
\displaystyle
\sum_{n=0}^N q_n \qty(\braket{\varphi_m}{\varphi''''_n} - 2 \alpha^2 \braket{\varphi_m}{\varphi''_n} + \alpha^4 \braket{\varphi_m}{\varphi_n}) \\
\displaystyle
\hspace{5.9cm}+\; i\, \frac{\alpha}{2}\,Re_E\,  \, \sum_{n=0}^N s_n \braket{\varphi_m}{f'\, \tilde\varphi_n} = 0 .
\end{cases}
}
Here, the scalar product $\braket{\cdot\,}{\cdot}$ is just meant to be an integration over $z \in [-1,1]$.
The linear algebraic system \eqref{B3} can be rewritten in terms of $(2N+2) \times (2N+2)$ block matrices,
\eq{B4}{
\mqty({\boldsymbol{\cal A}}^{(1)} & \vb{0} \\ \vb{0} & {\boldsymbol{\cal A}}^{(2)}) \mqty(\vb{s}\\ \vb{q}) = Re_E \, \mqty(\vb{0} & {\boldsymbol{\cal B}}^{(1)}\\ {\boldsymbol{\cal B}}^{(2)} & \vb{0}) \mqty(\vb{s}\\ \vb{q}) ,
}
where
\eq{B5}{
{\cal A}_{m n}^{(1)} = \braket{\tilde\varphi_m}{\tilde\varphi''_n} - \alpha^2 \braket{\tilde\varphi_m}{\tilde\varphi_n}, \nn\\
{\cal A}_{m n}^{(2)} = \braket{\varphi_m}{\varphi''''_n} - 2 \alpha^2 \braket{\varphi_m}{\varphi''_n} + \alpha^4 \braket{\varphi_m}{\varphi_n}, \nn\\
{\cal B}_{m n}^{(1)} =  - i\, \frac{\alpha}{2} \braket{\tilde\varphi_m}{f'\, \varphi_n}, \nn\\
{\cal B}_{m n}^{(2)} =  - i\, \frac{\alpha}{2} \braket{\varphi_m}{f'\, \tilde\varphi_n}. 
}
We note that ${\boldsymbol{\cal B}}^{(2)}$ is the transpose of ${\boldsymbol{\cal B}}^{(1)}$. The generalised eigenvalue problem defined by \eqref{B4}, where $Re_E$ is the eigenvalue, can be solved by employing the software {\sl Mathematica 14} ({\small \copyright}\ Wolfram Research, Inc.). We consider only the case of the Poiseuille flow, as the Couette flow differs just in the expression of $f(z)$. The first step is defining the matrix elements after fixing a maximum value for $N$, say $N=10$,

\begin{table}[t]
\centering
\begin{tabular}{|c|c|c|}
\hline
number of terms & $Re_E$ & $Re_E$\\
\hline\hline
$8+1$   & 49.62031330610141 & 61.74928015480069\\
$10+1$ & 49.62030652892564 & 61.74927602800918\\
$12+1$ & 49.62030651974471 & 61.74927601048756\\
$14+1$ & 49.62030651973459 & 61.74927601045481\\
$16+1$ & 49.62030651973426 & 61.74927601045474\\
$18+1$ & 49.62030651973426 & 61.74927601045474\\
$20+1$ & 49.62030651973426 & 61.74927601045474\\
\hline
\end{tabular}
\caption{\label{tabB1}\centering{}L-modes eigenvalue problem for the Poiseuille flow, with $\alpha=2$: \linebreak the smallest two positive eigenvalues $Re_E$, with increasing \linebreak number of terms in the series approximation}
\end{table}

\begin{footnotesize}
\begin{verbatim}
NN=10; f[z_]:=1-z^2; 
phi[n_,z_]:=(1-z^2)^2 ChebyshevT[n,z]; phit[n_,z_]:=(1-z^2) ChebyshevT[n,z];
Do[E0[m,n]=Integrate[phit[m,z] phit[n,z],{z,-1,1}],{m,0,NN},{n,0,NN}];
Do[E2[m,n]=Integrate[phit[m,z] D[phit[n,z],{z,2}],{z,-1,1}],{m,0,NN},{n,0,NN}];
Do[EU0[m,n]=Integrate[phit[m,z] f'[z] phi[n,z],{z,-1,1}],{m,0,NN},{n,0,NN}];
Do[A0[m,n]=Integrate[phi[m,z] phi[n,z],{z,-1,1}],{m,0,NN},{n,0,NN}];
Do[A2[m,n]=Integrate[phi[m,z] D[phi[n,z],{z,2}],{z,-1,1}],{m,0,NN},{n,0,NN}];
Do[A4[m,n]=Integrate[phi[m,z] D[phi[n,z],{z,4}],{z,-1,1}],{m,0,NN},{n,0,NN}];
Do[AU0[m,n]=Integrate[phi[m,z] f'[z] phit[n,z],{z,-1,1}],{m,0,NN},{n,0,NN}];
EE0[alpha_,M_]:=Table[E2[m,n]-alpha^2 E0[m,n],{m,0,NN},{n,0,NN}];
AA0[alpha_,M_]:=Table[A4[m,n]-2 alpha^2 A2[m,n]+alpha^4 A0[m,n],{m,0,NN},{n,0,NN}];
AA[alpha_,M_]:=ArrayFlatten[BlockDiagonalMatrix[{EE0[alpha,M],AA0[alpha,M]}]];
BB[alpha_,M_]:=ArrayFlatten[{{ConstantArray[0,{M+1,M+1}],
Table[-I alpha/2 EU0[m,n],{m,0,M},{n,0,M}]},
{Table[-I alpha/2 AU0[m,n],{m,0,M},{n,0,M}],ConstantArray[0,{M+1,M+1}]}}];
\end{verbatim}
\end{footnotesize}

\noindent{}If the value of $N$ is fixed, one can always test the accuracy of the approximation for every truncation order less or equal than $N$. For instance, one may obtain the eigenvalues for $\alpha = 2$ and a truncation to the first $8+1$ terms which, in this case, means square matrices of dimension $2(8+1)$,

\begin{footnotesize}
\begin{verbatim}
NumberForm[SortBy[With[{alpha=N[2,40],M=8},Chop[Eigenvalues[{AA[alpha,M],BB[alpha,M]}]]],
N[Abs[Re[#]]]&],6]
\end{verbatim}
\end{footnotesize}

\noindent{}The output for $Re_E$ is

\begin{footnotesize}
\begin{verbatim}
{-49.6203,49.6203,-61.7493,61.7493,-285.696,285.696,-342.821,342.821,-947.831,947.831,
-1092.80,1092.80,-3375.90,3375.90,-3934.38,3934.38,Infinity,Infinity}
\end{verbatim}
\end{footnotesize}

\noindent{}In this case, with $8+1$ terms in the series expansion, we get $18$ eigenvalues. The eigenvalues $|Re_E| = \infty$ may just be ignored. We note that the approximation of the smallest eigenvalue pair $\pm|Re_E|$ is good, while the accuracy for eigenvalues with a larger absolute value gradually decreases. A test of the convergence in the evaluation of the smallest two positive eigenvalues is reported in Table~\ref{tabB1}.

\end{appendices}

\end{document}